\documentclass[preprint,review,12pt]{elsarticle}

\usepackage{amssymb}

 \usepackage{lineno}

\bibpunct{(}{)}{;}{a}{}{,}

\journal{Icarus}

\usepackage{endfloat}
\usepackage{ulem}

  \newcommand{\degr}{\ensuremath{^\circ}}
  \newcommand{\arcsec}{\ensuremath{^{\prime\prime}}}

  \newcommand{\add}[1]{}
  \newcommand{\ben}[1]{}
  \newcommand{\boss}[1]{}

  \newcommand{\rr}[1]{}

\newcommand{\tiu}{\ensuremath{J~m^{-2}~s^{-0.5}~K^{-1}}}

\usepackage{hyperref}
\hypersetup{
    bookmarks=true,         
    unicode=false,          
    pdftoolbar=true,        
    pdfmenubar=true,        
    pdffitwindow=false,     
    pdftitle={Thermophysical modeling of (16) Psyche},
    pdfauthor={A. Matter et al.}, 
    pdfsubject={Planetary Science}, 
    pdfnewwindow=true,      
    colorlinks=true,        
    linkcolor=gray,         
    citecolor=blue,         
    filecolor=gray,         
    urlcolor=gray           
}

\begin{document}

\begin{frontmatter}



  \title{Evidence of a metal-rich surface for the asteroid (16) Psyche from interferometric observations in the thermal infrared
   \tnoteref{ESOID}} 
    \tnotetext[ESOID]{Based on observations collected
    at the European Southern Observatory (ESO), Chile: ESO Program ID 386.C-0928}


\author[MPIFR]{Alexis Matter\corref{cor1}\fnref{fn2}}
\cortext[cor1]{Corresponding author.}
\fntext[fn2]{Present address: Institut de plan\'{e}tologie et d'astrophysique de Grenoble, 414, rue de la Piscine, 38400 Saint Martin d'H\`{e}res, France.}
\ead{alexis.matter@obs.ujf-grenoble.fr} 
\author[Lagrange]{Marco Delbo}
\ead{delbo@oca.eu} 
\author[IMCCE]{Beno{\^i}t Carry}
\ead{bcarry@imcce.fr} 
\author[INAF]{Sebastiano Ligori}
\ead{ligori@oato.inaf.it} 

\address[MPIFR]{Max Planck institut f\"{u}r Radioastronomie, auf dem H\"{u}gel, 69, 53121, Bonn, Germany.}
\address[Lagrange]{UNS-CNRS-Observatoire de la C\^ote d'Azur,
  Laboratoire Lagrange, BP 4229 06304 Nice cedex 04, France.}
\address[IMCCE]{IMCCE, Observatoire de Paris, UPMC, CNRS, 77 Av. Denfert Rochereau, 75014 Paris, France.}  
\address[INAF]{INAF-Osservatorio Astronomico di Torino, Strada
  Osservatorio 20, 10025 Pino Torinese, Torino, Italy.}
\end{frontmatter}
\begin{flushleft}
\vspace{2cm}
Number of pages: 45\\
Number of tables: 2\\
Number of figures: 5\\
\end{flushleft}

\newpage
\begin{flushleft}
\begin{Large}
\vspace{1cm}
Proposed running head: Infrared interferometry of the asteroid (16) Psyche.\\

\vspace{1cm}
\noindent Editorial correspondence and proofs should be directed to: \\
Alexis Matter\\
 Institut de Plan\'{e}tologie et d'Astrophysique de Grenoble, \\
 414 rue de la Piscine, 38400 Saint Martin d'H\`{e}res, France.\\
email: alexis.matter@obs.ujf-grenoble.fr; \\
Tel: +33 4 76 63 58 30\\
Fax: +33 4 76 44 88 21
\end{Large}
\end{flushleft}

\clearpage
\section*{Abstract}
\linenumbers	
  
We describe the first determination of thermal properties and size of the
  M-type asteroid (16) Psyche from interferometric observations obtained with
  the Mid-Infrared 
  Interferometric Instrument (MIDI)
  of the Very Large
  Telescope Interferometer. We used a thermophysical model
  to interpret our  interferometric data.  
  Our analysis shows 
  that Psyche has a low macroscopic surface
  roughness.  Using a
  convex 3-D shape model obtained by Kaasalainen et al. (2002. Icarus 159,
  369--395), we derived a volume-equivalent diameter for (16) Psyche of
  $247\pm25$~km or $238\pm24$~km, depending on the possible values of surface roughness.  
  Our corresponding thermal inertia estimates are 133 or 114~\tiu, with a total uncertainty estimated at
  40~\tiu.
  They are among the highest thermal inertia values ever measured for an asteroid
  of this size. We consider this as a new evidence of a metal-rich
  surface for the asteroid (16) Psyche.


\vspace{3cm}
\section*{Keywords}
\noindent
  Asteroids; \sep Asteroids surfaces; \sep
  Infrared observations; \sep Data reduction techniques.

%

%


\newpage
\section{Introduction}

  \indent Asteroids classified in the X-complex \citep[
    in the taxonomies by
  ][]{2002Icar..158..146B,2009Icar..202..160D} are characterized by a visible
  and near-infrared reflectance spectrum that is essentially 
  featureless and moderately red in the [0.3-2.5] micron region.
  The spectroscopic X-complex can be split into 
  three taxonomic classes, E, M and P,
  according to albedo \citep{1984PhDT.........3T}.
  M-type asteroids are distinguished by exhibiting
  moderate geometric visible albedos of about 0.1 to 0.3.
  Due to the lack of  
  absorption features in the spectrum of M-type asteroids,
  the nature of these objects
  remains uncertain. Historically, M-class
  asteroids were assumed to be the exposed metallic core of differentiated
  parent bodies that were catastrophically disrupted, and thus the source of iron
  meteorites \citep{1989aste.conf..921B,1990JGR....95.8323C}. 
  While the parent bodies of meteorites are usually assumed to have formed in the main
  belt, \citet{2006Natur.439..821B} showed that the iron-meteorite
  parent bodies most probably formed in the terrestrial planet region. Some of
  the metallic objects currently located in the main-belt may thus not be
  not indegenous and possibly remnants of the precursor 
  material that formed the
  terrestrial planets including the Earth. Therefore, those objects play a
  fundamental role in the investigations of the solar system formation 
  theories.   
  \indent Radar observations provided strong evidences for the metallic
  composition of a least some M-type asteroids. Very high radar albedos have
  been measured for various asteroids of this class, consistent with high 
  concentration of metal \citep{1985Sci...229..442O,2008Icar..195..184S}. Moreover, the
  average density of two multiple M-type asteroids, 3.35~g.cm$^{-3}$
  \citep{2008Icar..196..578D} for (22) Kalliope and 3.6~g.cm$^{-3}$
  \citep{2011Icar..211.1022D} for (216) Kleopatra, appeared to be significantly
  larger than the density of C-type or S-type
  asteroids \citep{2012P&SS...73...98C}. This
  is a strong evidence of difference in internal composition between M and
  C-type asteroids. However, recent visible and near-infrared spectroscopic surveys on about 20 M-type
  asteroids, including those exhibiting high radar albedos, detected subtle
  spectral 
  absorption features on most of them
  \citep{2005Icar..175..141H,2010Icar..210..655F}. The most common one being the
  0.9~$\mu$m absorption feature, attributed to orthopyroxene, and thus
  indicating the presence of silicate on their surface. From a survey of the
  3~$\mu$m spectrum of about 30 M-type asteroids,
  \citet{1995Icar..117...90R,2000Icar..145..351R} also found hydration features
  on a tens 
  of them. On the basis of these observations, they suggested that
  the original ``M'' class should be divided into ``M'' asteroids that lack
  hydration features such as (16) Psyche and (216) Kleopatra, and ``W'' asteroids
  that are hydrated such as (21) Lutetia. 
  All of that confirms that most of the objects defined by the Tholen M-class
  have not a pure metallic surface composition but contain other species
  including silicate minerals.       
  Therefore, better compositional constraints for the spectrally 
  featureless bodies like
  M-type asteroids are essential in order to better understand and constrain
  the thermal, collisional, and migration history of Main-Belt Asteroids
  (MBAs). This includes the detection of additional absorption features in
  their reflectance spectra and the determination of their surface properties
  including surface roughness and in particular thermal inertia. 

  \indent Thermal inertia ($\Gamma$)
  is a measure of the resistance of a material to temperature
  change. It is defined by $\Gamma=\sqrt{\rho \kappa c}$, where $\kappa$ is
  the thermal conductivity, $\rho$ the material density and $c$ the 
  specific heat. The value of thermal inertia thus depends on the
  material properties \citep[see][and references therein for a table
    of the value of the thermal inertia of some typical
    materials]{Mueller2007}. On one hand, it 
  primarily informs us about the nature of the surface regolith: a soil with
  a very low value of $\Gamma$, for instance in the range between 20
  and 50 \tiu, is covered with fine dust
  like on Ceres \citep{1998A&A...338..340M}; an intermediate value
  (150-700 \tiu) indicates a coarser, mm- to cm-sized, regolith as
  observed on (433) Eros
  \citep{2001Natur.413..390V,2001Sci...292..484V} and (25143) Itokawa
  \citep{Yano2006}, respectively; solid rock with very little porosity
  is known to have thermal inertia values of more than 2500 \tiu\
  \citep{Jakosky1986}. On the other hand, thermal inertia  can represent a
  proxy for the surface composition, especially due to its dependency on
  thermal conductivity and specific heat. This is particularly important in
  the context of the M-type asteroids study since metal is an excellent
  thermal conductor, potentially leading to an enhanced thermal inertia. The
  study of \citet{2010Icar..208..449O} showed that thermal conductivity is
  significantly higher for iron meteorites than for non-metallic ones. This
  motivates our work of determining thermal inertia on M-type asteroids such
  as (16) Psyche to assess the change in thermal inertia for asteroids of
  different composition but having a similar size, knowing that the presence and thickness of the surface regolith is generally assumed to depend on the asteroid's size \citep[see, e.g.,][]{2005Icar..179...63B}.  
  
  \indent The asteroid (16) Psyche is the largest known M-type asteroid, with an IRAS
  diameter of $253\pm4$~km \citep{2002AJ....123.1056T}. Nevertheless, many
  size estimates have been reported during the last
  decade. \citet{2003Icar..162..278C} derived an area equivalent diameter of
  $288\pm43$~km based on speckle interferometry; \citet{2006SoSyR..40..214L}
  derived a diameter of 213~km based on considerations on its polarimetric
  albedo;
  from adaptive-optics imaging, \citet{2008Icar..197..480D} derived a volume equivalent
  diameter of $262\pm6$~km; \citet{2008Icar..195..184S} derived a volume equivalent diameter
  of $186\pm30$~km based on radar imaging; 
  from the analysis of medium infrared data from the AKARI
    satellite by means of the Standard Thermal Model \citep{1986Icar...68..239L}, \citet{2011PASJ...63.1117U} derived a diameter of $207\pm3$~km; finally,
  \citet{2011Icar..214..652D} derived a volume equivalent diameter of
  $211\pm21$~km by combining a shape model derived by lightcurve inversion
  with occultation observations of (16) Psyche. In any case, Psyche
  appears to be significantly larger than the 30-90~km diameter expected for
  the metallic core of a differentiated asteroid
  \citep{2000Icar..145..351R}, questioning a purely metallic nature for this asteroid. 
  All those size measurements led to significant differences between the
  average bulk density estimations reported in the literature. They range from $1.8\pm0.6$~g.cm$^{-3}$
  \citep{2000A&A...354..725V} to  $3.3\pm0.7$~g.cm$^{-3}$
  \citep{2006DPS....38.5925D} and even $6.58\pm0.58$~g.cm$^{-3}$
  \citep{2002A&A...395L..17K}, value which is more in agreement with a
  metallic composition and a very low macroporosity. 
  Nevertheless, by combining all the
  independent size and mass estimates, an 
  average density of $3.36\pm1.16$~g.cm$^{-3}$ was found
  \citep{2012P&SS...73...98C}. This is comparable to the density estimates reported for other M-type asteroids like (22) Kalliope \citep{2008Icar..196..578D} and (216) Kleopatra \citep{2011Icar..211.1022D}. In addition, \citep{2010Icar..208..221S} measured a high radar albedo of 0.42, which is indicative of a metal-rich surface. However, the detection of a
  0.9~$\mu$m absorption feature suggested the presence of silicates on its
  surface \citep{2005Icar..175..141H}. In this context, \citet{2005Icar..175..141H} and \citet{2010Icar..208..221S} suggested that (16) Psyche may be a collisional aggregate of several objects, including partial
  or intact metallic cores that have retained a portion of their silicate-rich. 

  \indent To put tighter constrains on the nature of (16) Psyche, we used
  mid-infrared interferometry to determine the thermal
  properties of this asteroid, and refine its size
  measurements. Interferometry basically provides direct measurements of the
  angular extension of the asteroid along different directions
  \citep{2009ApJ...694.1228D}. Interferometric asteroid observations in the thermal infrared, where 
  the measured flux is dominated by the body's thermal emission, are 
  sensitive to the surface temperature spatial distribution in different
  directions on the plane of the sky. The typical spatial resolution is about 0.06~\arcsec in the case of our Psyche observations. 
  As the surface temperature distribution
  of atmosphereless bodies is affected by thermal inertia and surface
  roughness, interferomeric thermal infrared data can be used to constrain these parameters. In particular, thermal
  infrared interferometry can help to remove
  the degeneracy existing between the effect of the thermal inertia and surface
  roughness in one single epoch \citep[see Figs.~7 and~8 in][]{2011Icar..215...47M}, providing that we have several interferometric measurements with different projected baseline lengths and orientations, during the asteroid rotation.
  Thermal properties 
 (thermal inertia and surface roughness) can thus be better constrained by thermal
 infrared interferometry in combination with the classical disk-integrated
 radiometry. In this context, we obtained interferometric data on (16) Psyche
 using the MIDI instrument combining two of the Auxiliary Telescopes (ATs) of
 the 
 VLTI. As in  \citet{2011Icar..215...47M}, a thermophysical model (TPM),
   taking into account the asteroid's orbit, spin, shape, and heat diffusion
   into the subsurface, was
 used for the analysis of the whole data set. 

 \indent In section~\ref{s:MIDI}
 we report the observations and the data reduction process that we
 adopted; in section~\ref{s:models} we briefly remind the principles of the
 thermophysical model used for the interpretation of MIDI 
 data, and we detail the shape models that we used; in section~\ref{s:results},
 we present our results, followed 
 by a discussion in section~\ref{s:discussion}.

\section{Observations and data reduction}
\label{s:MIDI}
\subsection{Observations}
  The observations of (16) Psyche were carried out in visitor mode, on 2010
  December 30. 
  Two ATs were used in the E0-G0 configuration (baseline $B=16$~m). Sky
  quality was relatively good and stable during those nights (see
  Table~\ref{tab:log}). We adopted the typical observing sequence of MIDI,
  which is extensively described by \citet{2004A&A...423..537L}. For each of
  the five observing epochs of (16) Psyche (indicated in Table~\ref{tab:log}),
  we obtained one measurement of the total flux and of the interferometric
  visibility, both dispersed over the N-band, from 8 to 13~$\mu$m. We
  used the HIGH-SENS mode, where the total flux of the source is measured right
  after the fringe tracking and not simultaneously. To disperse the fringes,
  we used the prism of MIDI, which gives a spectral resolution of
  $\frac{\lambda}{\Delta \lambda}\approx 30$ at $\lambda=10$~$\mu$m. Our
  observations also included a mid-infrared photometric and 
  interferometric calibrator HD~29139, taken from the ESO database
  using the Calvin tool \footnote{Available at
    http://www.eso.org/observing/etc/}, which is the calibrator selector for
  the VLTI instruments (MIDI and AMBER). We remind that interferometric
  calibrators are stars that have small 
  and known angular diameter, so that their visibility is close to
  unity at all wavelengths. This calibrator was chosen to have a
  minimum angular separation with the source ($\approx3$~\degr) and a similar
  airmass, as shown in Table~\ref{tab:log}. This Table also summarizes the log
  of observations, with the corresponding interferometric parameters.\\

 
 \begin{table}[h]
 \centering
 \vspace{3mm}
 \begin{tabular}[!h]{ccccccc}
 \hline
 Object& Date& PBL& PBLA & seeing& airmass& Label\\
& (UT) & (m) & (\degr) & (\arcsec) \\
 \hline
     {\small HD 29139}   &2010-12-30 00:24& 11.3 & 82.0 & 1.50 & 1.70&{\small Calib}\\
     {\small (16) Psyche}&2010-12-30 00:48& 11.88& 82.7 & 1.15 & 1.60& 1\\
     {\small HD 29139}   &2010-12-30 02:59& 14.4 & 79.0 & 1.00 & 1.40&{\small Calib}\\
     {\small (16) Psyche}&2010-12-30 03:18& 15.8 & 74.1 & 0.70 & 1.37&2\\
     {\small (16) Psyche}&2010-12-30 03:30& 16.0 & 73.3 & 0.75 & 1.37&3\\
     {\small (16) Psyche}&2010-12-30 03:40& 16.0 & 72.6 & 0.70 & 1.39&4\\
     {\small HD 29139}   &2010-12-30 03:58& 15.9 & 70.5 & 0.60 & 1.40&{\small Calib}\\
     {\small HD 29139}   &2010-12-30 04:33& 15.6 & 67.5 & 0.64 & 1.50&{\small Calib}\\
     {\small (16) Psyche}&2010-12-30 04:51& 15.5 & 66.5 & 0.75 & 1.60&5\\
     \hline
 \end{tabular}
 \caption[Observation log]{Log of the observations for (16) Psyche and its calibrator HD 29139,
   both observed with two ATs in E0-G0 configuration. PBL and PBLA stand for
   Projected BaseLine, and Projected BaseLine Angle, respectively. The last
   column gives a label for each interferometric observation of (16) Psyche,
   the label `Calib' indicating a calibrator observation.} 
 \label{tab:log}

\end{table}
 
\subsection{Data reduction}
\begin{figure}
 \centering
 \includegraphics[width=130mm,height=115mm]{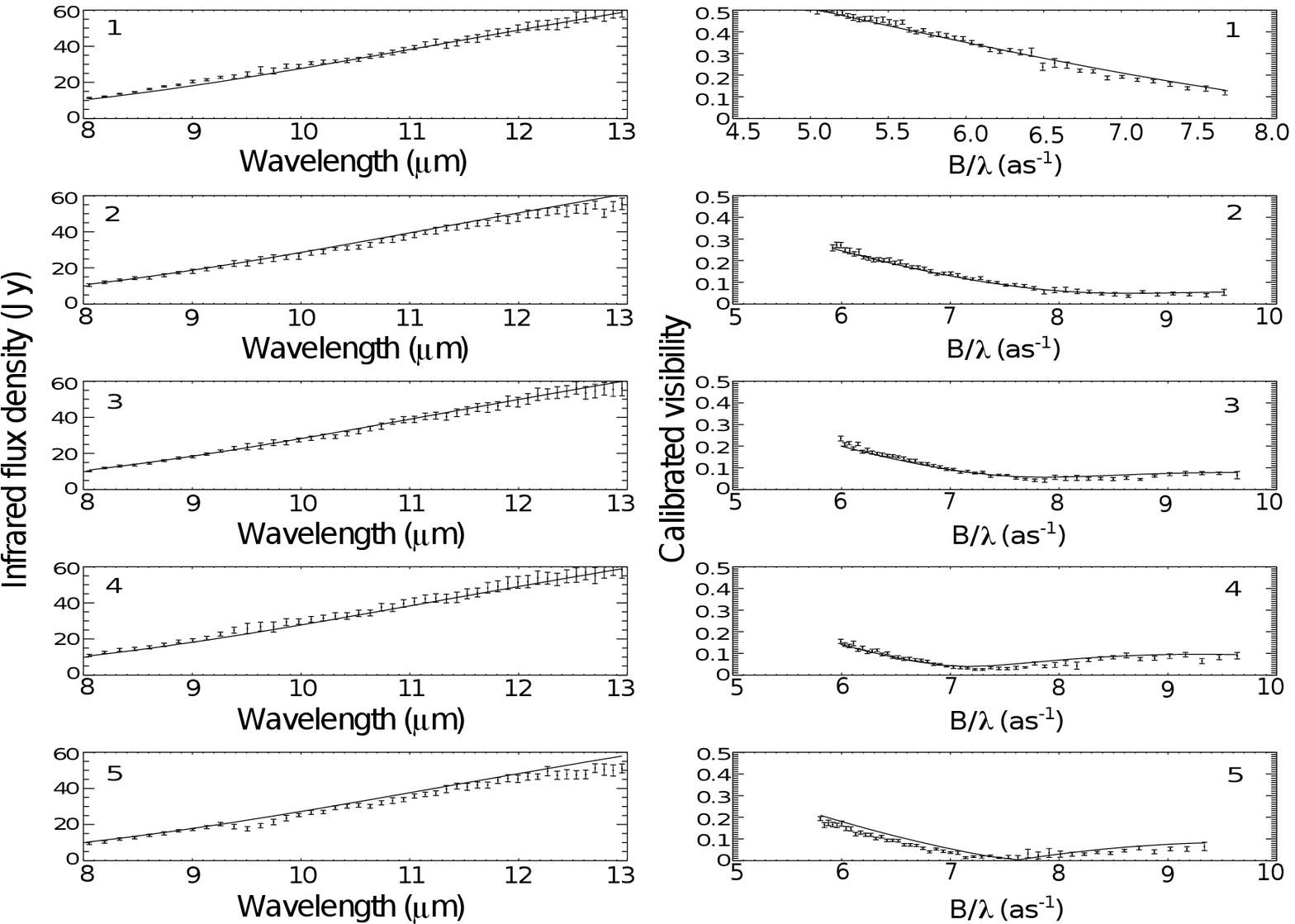}
 \caption[Observations]{
   \textbf{Left panels:} measured thermal infrared fluxes (with error bars) and best-fit TPM infrared fluxes (solid lines)
   of (16) Psyche, plotted between 8 and 13~$\mu$m.
   \textbf{Right panels:}
   measured mid-infrared interferometric visibilities (with error bars) and best-fit TPM visibilities 
   (solid lines) of (16) Psyche, plotted as a function of
   angular frequency. The best-fit model represented here is : `low roughness',
   $\Gamma=115$~\tiu. For each pair of flux and visibility measurements, we
   indicated the label (`1', `2', `3', `4', `5') of the corresponding
   observing epoch, as defined in Table~\ref{tab:log}.} 
 \label{fig:flux_vis_bestfit}
\end{figure}

  Extraction and calibration of the flux and visibility measurements of (16)
  Psyche were performed using the data reduction software package EWS (Expert
  WorkStation). This publicly available\footnote{Software package is available at
    http://home.strw.leidenuniv.nl/\~jaffe/ews/index.html}
  software performs a coherent analysis of dispersed fringes
  to estimate the complex visibility of the source. The method and the different
  processing steps are described in \citet{2004SPIE.5491..715J}. The calibration of
  the fluxes and visibilities was performed using the closest calibrator
  observation in time. Calibrated fluxes of (16) Psyche were
obtained by multiplying the ratio target/calibration star raw
counts, measured by MIDI at each wavelength, by the absolute flux of the calibrator. The absolutely calibrated infrared spectrum of our calibrator was taken from a database created by R. van Boekel, which is initially base on infrared templates created by \citet{1999AJ....117.1864C}. More details can be found in \citet{2005PhDT.........2V}. The instrumental visibilities of (16) Psyche correspond to the
ratio of the source raw correlated flux and the source raw photometric
flux. The calibrated visibilities of (16) Psyche were then derived by dividing
the instrumental visibility by the visibility measured on the calibrator for each observing epoch. We refer the reader to \citet{2009ApJ...694.1228D} for a more detailed description of the data reduction and calibration of MIDI interferometric data.\\ 
The five calibrated flux and visibility measurements of
  (16) Psyche are shown in 
  Fig~\ref{fig:flux_vis_bestfit}. The error bars represent the statistical noise contribution
  affecting the correlated and total flux measurements. It is estimated by
  splitting a complete exposure, consisting of several thousand of frames, into
  five equal parts and deriving the variance of these sub-observations. In the
  error budget, we neglected the error due to the variability of the
  interferometer transfer function during the night. However this is not a
  problem since, by computing the rms of the transfer function, i.e., the
  instrumental visibility provided by the calibrator observations, we found a
  very good stability all over the night.
\subsection{Observational results}
\begin{figure}
 \centering
 \includegraphics[scale=0.8]{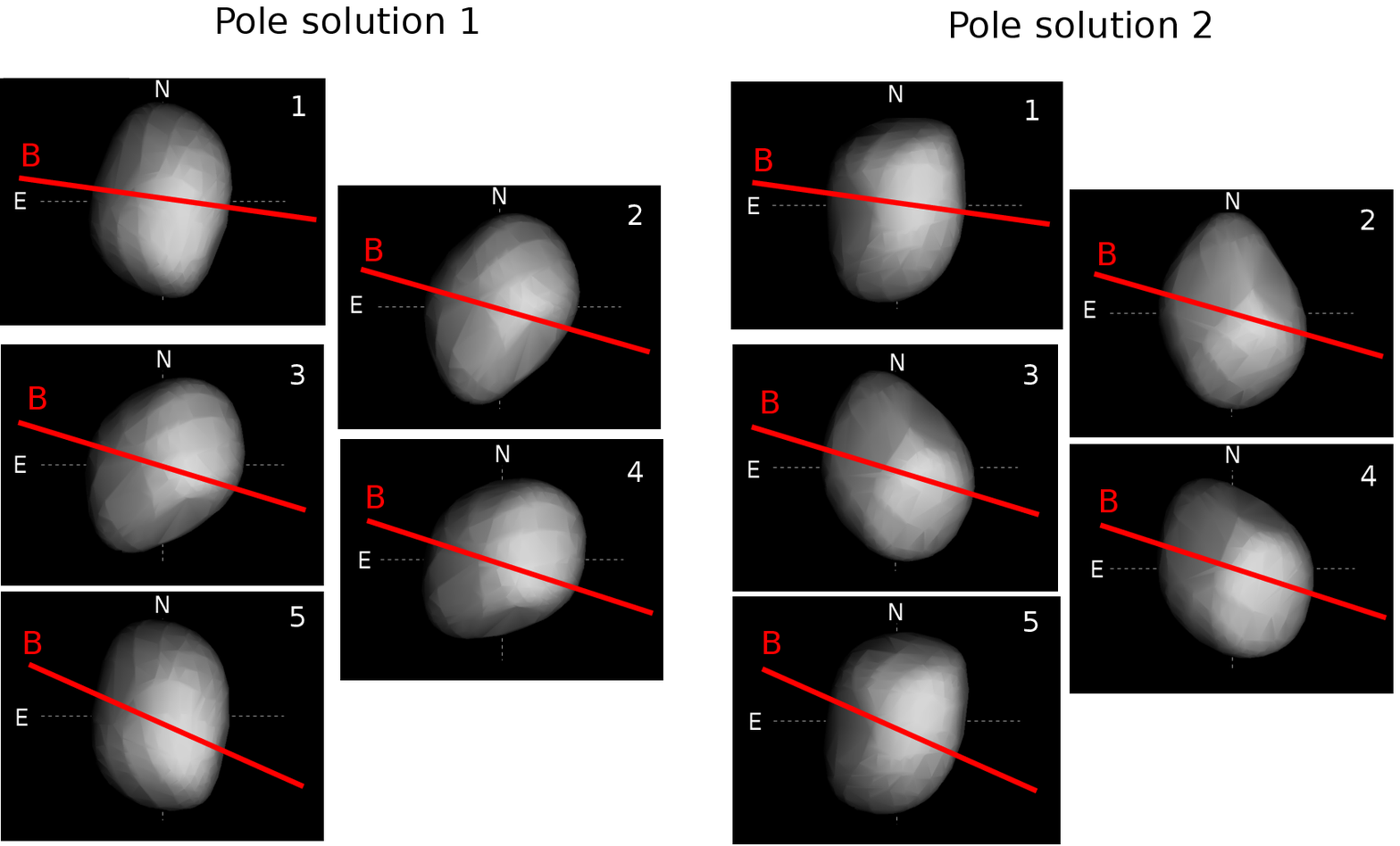}
 \caption[Predicted orientation of Psyche]{Representation of the expected orientation of (16) Psyche along with
   the projected baseline orientation at the time of the VLTI observations, on
   the plane of the sky; the North (N) and East (E) directions are indicated. We used the
   convex mesh and the two existing pole solutions derived by
   \citet{2002Icar..159..369K} and detailed in Section~\ref{s:models}. For
   each epoch, we indicated the corresponding label (1 to 5),
   as defined in Table~\ref{tab:log}.}
 \label{fig:VLTI_epochs}
\end{figure}

  At each observing epoch, the corresponding visibility is pretty low ($\approx
  0.1-0.4$), thus indicating that the object is very well resolved by MIDI. Moreover
  we can notice a significant decrease of the visibility level between the first
  and second epochs, which are separated by 2h30, and then only slight changes in the
  visibility shape and level for the next epochs. This behaviour can be
  explained by the increase of the length of the projected baseline during the
  observing night (see Table~\ref{tab:log}), combined to the evolution of the
  projected baseline orientation and the asteroid rotation. For instance, the
  difference between the first and last visibility measurements is mainly due to
  the increase of the projected baseline length since our observations covered a
  complete rotation of the asteroid ($\approx 4.2$~h), without a significant
  evolution of the baseline orientation. This is illustrated in
  Figure~\ref{fig:VLTI_epochs}, where we represented the expected orientation of
  (16) Psyche at the time of the VLTI observations. Here, we considered
  the two existing pole solutions derived by \citet{2002Icar..159..369K} and
  that we detail hereafter in Section~\ref{s:models}.

\section{Thermophysical modeling}
\label{s:models}

  \indent The thermophysical modeling of interferometric data is extensively
  described in \citet[and references therein]{2011Icar..215...47M}, and here
  we briefly remind its principles, along with the shape model we used for
  this modeling.\\ 
  The thermal inertia and macroscopic roughness of an asteroid can be derived by
  comparing measurements of the thermal-infrared flux and interferometric
  visibility of the body to synthetic fluxes and visibilities generated by means
  of a thermophysical model (TPM).  
  A TPM uses the spin vector information to orient a 3-D shape model, composed of a 
  mesh of planar facets, at the time of the observing epochs. The temperature of
  each facet is calculated by solving the one-dimensional heat diffusion
  equation (where the heat diffusion is between the surface and the shallow
  subsurface) using presets thermal inertia values. 
  Surface roughness is modeled
  by adding hemispherical craters of variable opening angle, $\gamma_c$, and
  surface density, $\rho_c$. As for \citet{2009P&SS...57..259D}, thermal
  conduction within the craters is also modeled. Albedo, thermal inertia, surface roughness and emissivity are assumed constant over the asteroid's surface. We remind that emissivity, noted hereafter $\epsilon$, has normally directional properties and drops at higher viewing angles, in particular at the limb of objects having a smooth surface \citep[see e.g.,][]{1990GeoRL..17..985J}. 
  However, we expect it to be a second-order effect since the limb normally contributes much less to the thermal emission than the nadir, especially if the object was observed under a low solar phase angle like for our observations of (16) Psyche. We also assumed $\epsilon$ achromatic and equal to 0.9, which is a typical value for silicate powders and is commonly assumed  for the surface of asteroids \citep{2006Icar..182..496E,Mueller2007}. However, regarding the possible metallic nature of the Psyche's regolith, it is worth mentioning that metals have usually emissivities lower than 0.9 depending on their state of roughness and porosity. For instance, powdered iron at a temperature of 300~K and with a porosity similar to the lunar regolith ($\approx 50$\%), has $\epsilon\approx0.8$ \citep{doi:10.1080/02726350490501682}. Assuming that the regolith of (16) Psyche is purely ferrous, we thus ran in parallel our TPM with $\epsilon=0.8$. As a result, only tiny changes were observed in terms of size, thermal inertia and surface roughness. Moreover, silicate material was detected on the Psyche's surface \citep{2005Icar..175..141H}, and could increase the surface emissivity. Therefore, we finally kept $\epsilon=0.9$ for the thermophysical modeling of (16) Psyche.\\   
Following the
  procedure described in \citet{2011Icar..215...47M}, the best-fit value of $a$,
  which is the linear mesh scale factor, for each discrete value 
  of $\Gamma$ and each roughness model can be found by
  minimizing a $\chi^2$ function taking into account both the integrated flux
  and the interferometric visibility. Then, the location of the minimum
  $\chi^2$ as a function of $\Gamma$ gives the best-fit 
  asteroid surface thermal inertia for each roughness model. Eventually, 
  the value of $a$ at $\Gamma$-minimum is used to determine the best-fit
  value of the volume equivalent diameter of the mesh, $D_\vee=2 \left( \frac{3
    \vee}{4 \pi} \right)^{\frac{1}{3}}$, where $\vee$ is the volume of the
  mesh.\\ 

  \indent Two convex mesh were downloaded from the Database of
  Asteroid Models from Inversion Techniques\footnote{
    \url{http://astro.troja.mff.cuni.cz/projects/asteroids3D/}}
    \citep[DAMIT, see][]{2010A&A...513A..46A}.
    Both
  shape models are characterized by a
  sidereal rotation period (P), and a pole solution giving the spin axis direction,
  which was initially derived
  by \citet{2002Icar..159..369K} from inversion of optical lighcurves. Note
  that while inversion of optical lightcurves reconstruct the 3D
  shape of an asteroid,
  these shapes are convex by construction and do not provide any size
  information. The models derived from the 
  optical lightcurves inversion are thus scaled to unity volume. Later on, the two shape
  models of (16) Psyche were refined and scaled by \citet{2011Icar..214..652D}
  using occultation data. The associated two possible pole solutions are:  
  \begin{itemize}
  \item solution 1: $\lambda_0$=32\degr~$\beta_0$=-7\degr,
    $P$=4.195948 h
  \item solution 2: $\lambda_0$=213\degr~$\beta_0$=0\degr,
    $P$=4.195948 h
  \end{itemize}
  where $\lambda_0$ and $\beta_0$ are the ecliptic longitude and latitude of the
  spin axis direction (J2000.0, in degree), and $P$ is the sidereal rotation period.
  The corresponding volume-equivalent diameters, derived from
  occultation data, are $211\pm21$~km (solution 1) and $209\pm29$~km (solution
  2). Interestingly, \citet{2011Icar..214..652D} reported that only the first
  pole solution reported by \citet{2002Icar..159..369K} was consistent with
  all the occultation chords. Moreover,
  \citet{2012DPS....4430208H} identified the same best pole solution from
  comparison of adaptive-optics images and the shape models. They derived an
  equivalent diameter of $209\pm9$~km. In parallel, \citet{2002Icar..159..369K} suggested the existence of albedo variegations over the surface of (16) Psyche. They detected the signature of a bright spot of moderate size, which is about 30\% brighter than the rest of the surface. However, albedo variegations have little effect on the thermal emission. In particular, by increasing the geometric visible albedo of 30\%, the 10~$\mu$m flux vary by about 2\%, which is within the error bars of our MIDI measurements. Therefore, the assumption of thermal homogeneity (thermal inertia, surface roughness) over the asteroid surface should not be affected by this bright spot.     
  
  Using both shape models, the TPM was run for each roughness model, and
  thermal inertia values of 5, 10, 
  25, 50, 75, 100, 125, 150, 175, 200, 300, 400, 500, 750 and 1000 \tiu. The
  roughness models we used are: `no roughness' ($\gamma_c=0^o$, $\rho_c=0$),
  `low roughness' ($\gamma_c=45^o$, $\rho_c=0.5$), `medium roughness'
  ($\gamma_c=68^o$, $\rho_c=0.75$), and `high roughness' ($\gamma_c=90^o$,
  $\rho_c=1.0$).     
  Then the fit procedure described in section~\ref{s:models} was applied to
  the measured fluxes and visibilities, each of them containing 47 points
  between 8 and 13~$\mu$m. The flux and visibility measurements shown in
  Figure~\ref{fig:flux_vis_bestfit} are the inputs of the thermophysical
  model. 

  In the next section we describe and discuss the results obtained
  from the application of the TPM to the observed visibilities and
  fluxes of (16) Psyche.

\section{Results}
\label{s:results}

\begin{figure}
\centering
 \includegraphics[width=65mm,height=48mm]{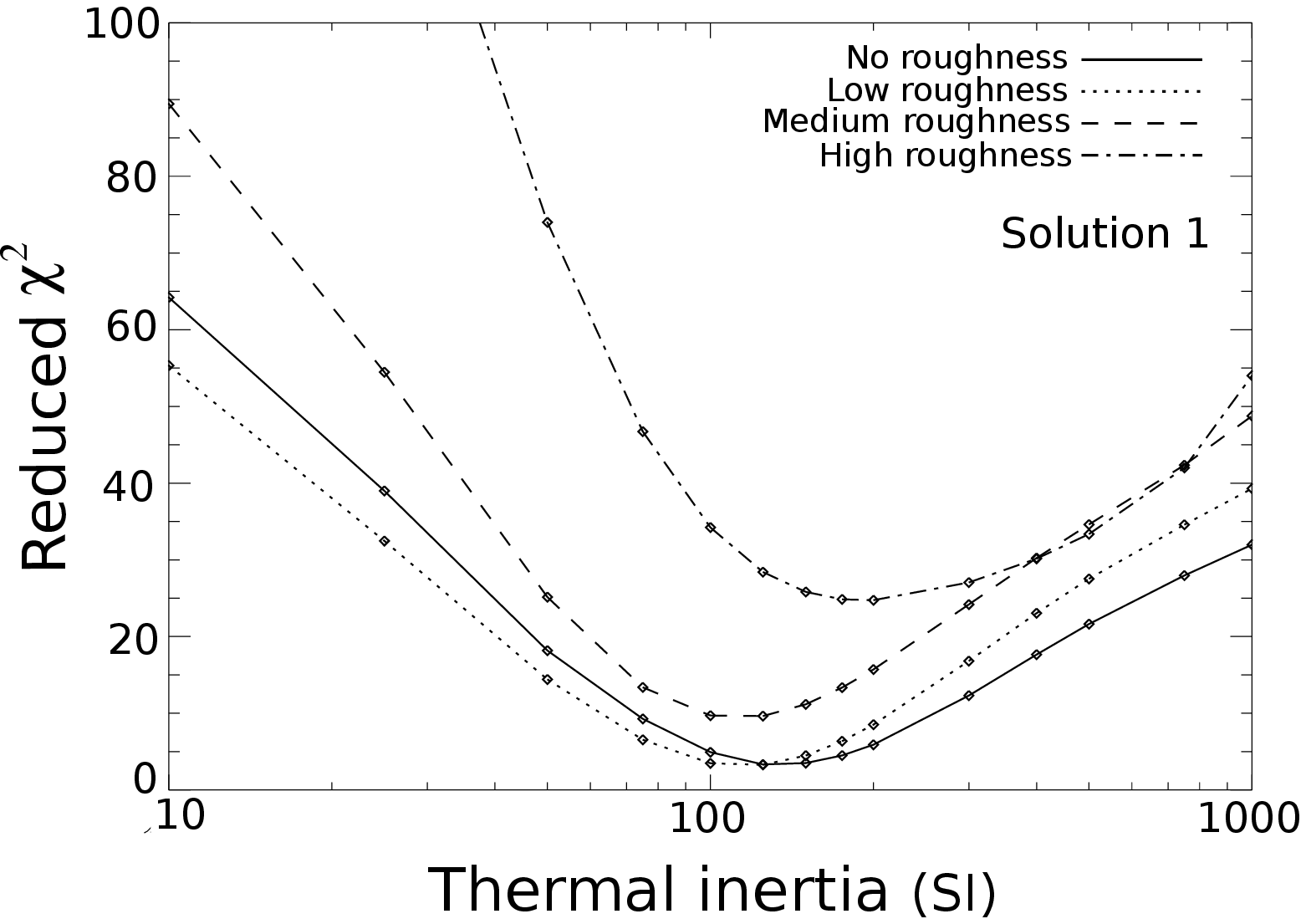}\hfill
 \includegraphics[width=65mm,height=48mm]{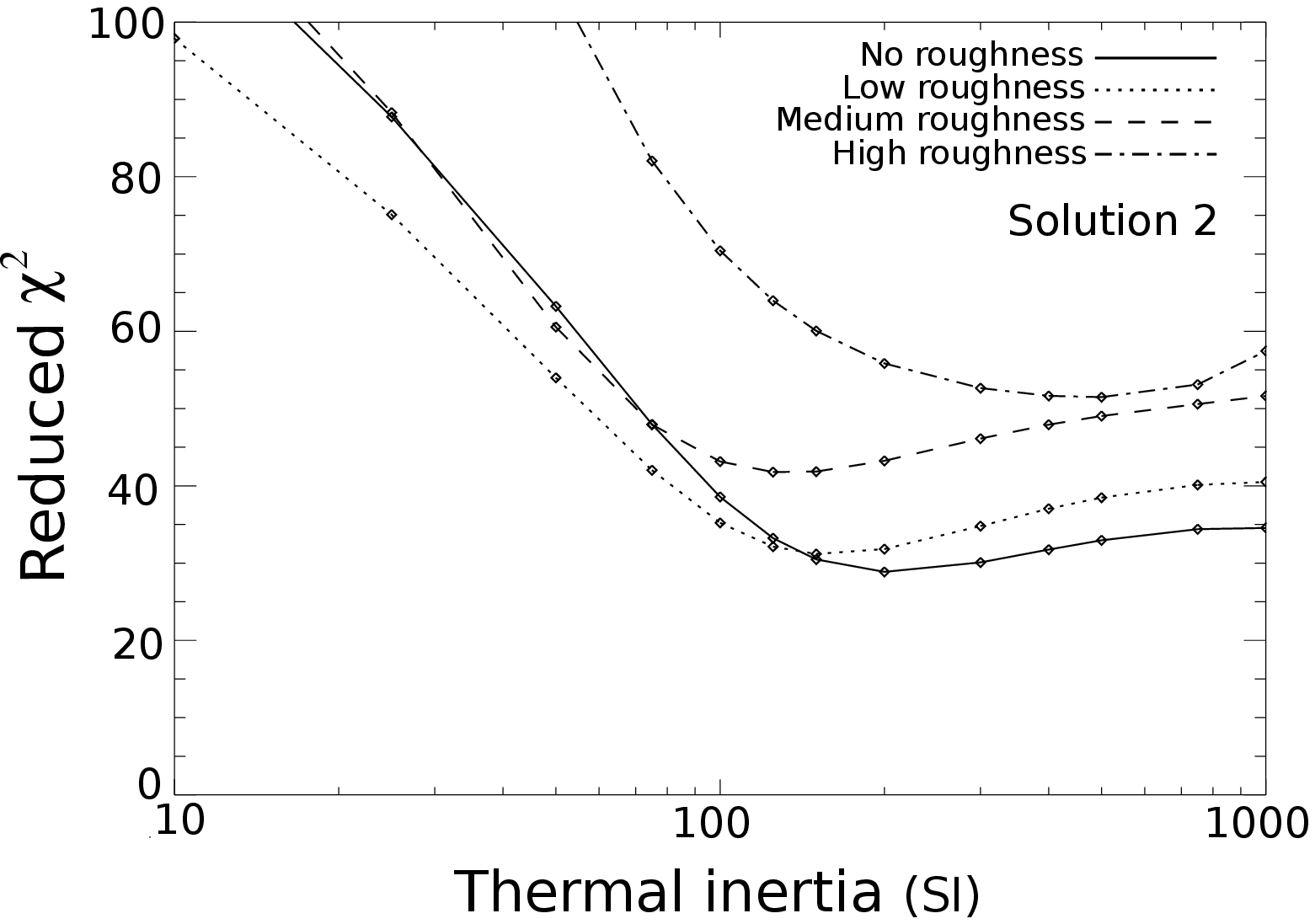}
 \caption[Thermal inertia determination]{
   \textbf{Left:} plot of the reduced $\chi^2$, calculated from the TPM in
   the case of the first pole solution, as a function of thermal inertia
   $\Gamma$, for the four roughness models (see Section~\ref{s:models}).
   \textbf{Right:} reduced $\chi^2$, calculated from the TPM in the case of the second
   pole solution. }
 \label{fig:chisq}
\end{figure}

  Fig.~\ref{fig:chisq} shows our reduced
  ${\chi}^2$ estimator as a function
  of $\Gamma$ for the four different roughness models, 
  in the case of the two pole solutions described above.\\

  We note that a surface with a low or no macroscopic roughness and a value of
  thermal inertia of about $130$~\tiu\ give the best fit to the observations,
  for both pole solutions. However, it clearly appears that the pole and
  shape solution 1 gives a better fit to our MIDI data than the pole and shape
  solution 2, by a factor of about 5.
  This is in agreement with the
  conclusions of \citet{2011Icar..214..652D} and \citet{2012DPS....4430208H} who
  favored the first pole solution. Therefore, we adopt the pole and shape
  solution 1 for the determination of thermal properties of (16) Psyche.   
  
\begin{figure}
\centering
 \includegraphics[scale=0.7]{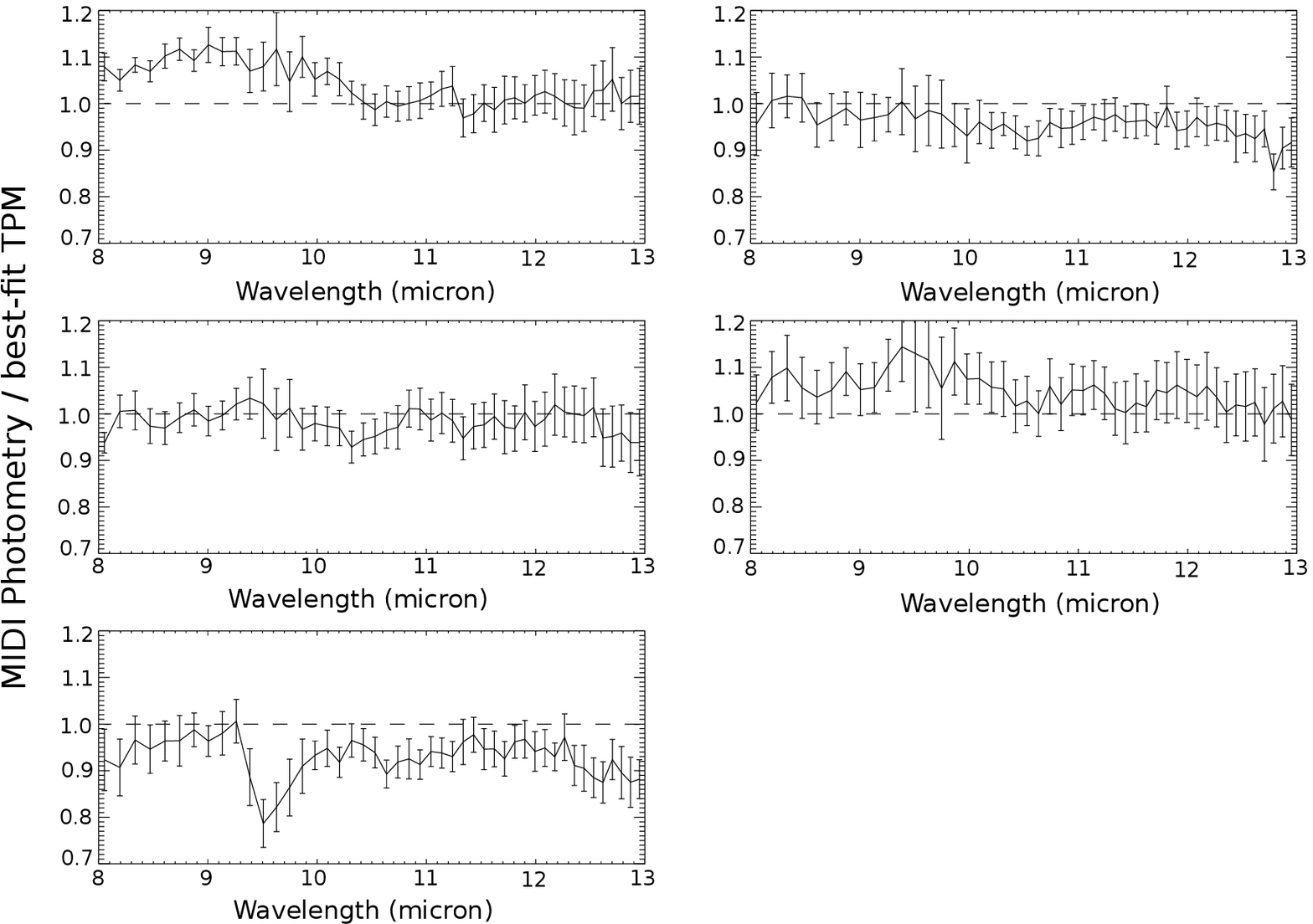}
 \caption[Ratio photometry/model]{Ratio of the MIDI flux over the best-fit TPM shown in Figure~\ref{fig:flux_vis_bestfit}, namely `low roughness' and $\Gamma=115$~\tiu. We indicated the label corresponding to each observing epoch (`1' to `5'). Since we assumed an emissivity of 0.9 in our TPM, the level of the continuum (dashed line) is at 1.
   }
 \label{fig:ratio}
\end{figure}

  \indent To refine the estimation of the best-fit thermal inertia values, we
  ran the TPM for additional values, namely, $\Gamma=105$, 110, 115, 120, 130,
  135, 140, 145, 155, 
  165. Then, using a polynomial interpolation, we found the minima of 
  the `no roughness' and `low roughness' models at $\Gamma$=133 and
  114~\tiu, respectively. The reduced $\chi^2$ values of those two best-fit
  models are 3.26 and 3.14, respectively. To obtain the best-fit values for
  $D_\vee$, we multiplied the volume-equivalent diameter of the mesh (211~km)
  by the best-fit scale factor values we got from our fit, namely 1.17 and
  1.13 for the `no roughness' and `low roughness' models, respectively. The
  corresponding values of $D_\vee$ are 247 
  and 238~km, with associated $p_V$ values of 0.12 and 0.13,
  respectively; for the calculation of $p_V$, we considered an absolute
  magnitude H of 5.93 according to \citet{2012Icar..221..365P}. In parallel, 
  we also forced the volume equivalent diameter of
  the mesh to its nominal value (211~km for the solution 1) in order to
  assess its effect on our thermal inertia determination.  We thus fixed the
  mesh scale factor to 1, and recalculated a $\chi^2$ value for each
  roughness model and thermal inertia value. It appeared that the fit to the
  visibility and flux measurements is significantly worse for any roughness
  model ($\chi^2_{\rm red} \geq 50$) while the best-fit values for thermal
  inertia are slightly lower (between 90 and 110~\tiu).    
  In Fig.~\ref{fig:flux_vis_bestfit} we plot the visibility
  and flux of the best-fit model (`low roughness', $\Gamma=114$~\tiu,
  $D_\vee=238$~km, $p_V=0.13$), in addition to the measured fluxes
  and visibilities of (16) Psyche. We note that our model represents
  well the observed flux except for the long-wavelength edge of the N band. Indeed, for the second and fifth observing epochs,
  the TPM flux is greater than the measured one by roughly
  10\% at 13~$\mu$m.  This `offset-like' mismatch may come
  from an underestimation of the total flux of the source by MIDI, due to a
  bad estimation and subtraction of the thermal background (and its
  fluctuations), which is dominant in 
  the mid-infrared \citep[see, e.g.,][]{2003EAS.....6..127P}. This is
  especially problematic around 13~$\mu$m where the atmospheric transmission
  starts to be degraded by water absorption lines.    
  Since the MIDI correlated flux measurements are usually not very much affected by background subtraction \citep[see e.g.,][]{2007NewAR..51..666C},
  an underestimation of the photometry would bring an increase in the
  visibility. However, since the fringe contrast is low, this effect is not
  noticeable in our visibility measurements. The fit to the visibilities
  generally appears good and follow the same trend as the measurements,
  confirming the good match of the shape model to the interferometric data.\\
%
\indent In order to search for the presence of possible emission features in the mid-infrared spectrum of (16) Psyche, we plotted in Figure~\ref{fig:ratio} the ratio between the MIDI flux measurement and our best-fit TPM for each observing epoch. Only the first epoch shows marginal detection of a possible emission feature between 8 and 10~$\mu$m that could be associated with the Christiansen peak around 9 um. The error bars shown in Figure~\ref{fig:ratio} represent the statistical uncertainty affecting the MIDI measurements. However, bad estimation and removal of the strong thermal background over the N band and/or the atmospheric ozone absorption feature around 9.6~$\mu$m can imply additional uncertainties of the order of 10\% on the absolute level of MIDI photometry measurements \citep[see e.g.,][]{2007NewAR..51..666C}. We thus think this most likely applies to the `absorption feature' around 9.6~$\mu$m in the fifth epoch plot, and then probably also to the `emission' pattern between 8 and 10~$\mu$m in the first epoch plot. As a consequence, we cannot report with confidence the detection of an emission feature in the mid-infrared spectrum of (16) Psyche. 
  
  To estimate the statistical uncertainty affecting the fit
  parameters $\Gamma$ and $D_\vee$, we followed the Monte-Carlo procedure
  described in \citet{2011Icar..215...47M}: 200 normally distributed flux and
  visibility values 
  per observation were generated at each wavelength, with average and standard
  deviation matching the data within their respective 1-$\sigma$ uncertainty;
  then for each set of fluxes and visibilities, a new $\chi^2$ was computed;
  finally we took the standard deviation of the best-fit $\Gamma$ and $D_\vee$
  values, found for all the synthetic data set, as the 1-$\sigma$ uncertainty
  on our best-fit value for $\Gamma$ and $D_\vee$. As a result we find 
  $\Gamma=133\pm2$~\tiu, $D_\vee=247\pm1$~km as the best fit
  solution for a model without roughness; and
  $\Gamma=115\pm2$~\tiu, $D_\vee=238\pm1$ km as the best fit
  solution for a model with a low roughness. Therefore, the 1-$\sigma$
  statistical error from the Monte Carlo procedure is about 1\% on the volume
  equivalent diameter and 2\% on the thermal inertia. This is very low, as
  expected from the statistical noise affecting the MIDI measurements, and
  probably underestimates the true uncertainty on our thermal inertia and diameter
  estimates. Indeed, in thermophysical modeling, this uncertainty is generally
  dominated by the model systematics, as this was the case for the first thermophysical modeling of the interferometric data of (41) Daphne. \citet{2011Icar..215...47M} estimated such a systematic error to be of about 7\% on the diameter, taking into account only the contribution from the surface roughness modeling.  
In the case of (16) Psyche, the model systematics due to the surface roughness modeling is
estimated by considering the two plausible solutions in terms of our best-fit 
  indicator, i.e., no roughness and $\Gamma$=133~\tiu, and low roughness and
  $\Gamma$=114~\tiu. The corresponding uncertainty would thus be 9
  km, i.e., 3\% in 
  relative uncertainty for $D_\vee$, and 19~\tiu, i.e., 15\% in relative
  uncertainty for $\Gamma$. However, the systematic uncertainty budget probably includes additional
  contributions from the spin solution, the assumption of isotropic emissivity, the shape model itself and especially the albedo variegations that impact the optical lightcurves inversion process. This last point may be important
  here since \citet{2002Icar..159..369K} found evidences of a bright spot (about 30\% brighter) on one side of the (16) Psyche's surface. However, we could not properly estimate the
  contribution of these sources of uncertainty in the TPM. Therefore, we finally
  adopted a conservative error value of 10\% in $D_\vee$, as generally
  considered in other thermophysical models \citep[see,
    e.g.,][]{2010Icar..205..505M,2012Icar..221.1130M}. For thermal inertia, we
  estimated a conservative uncertainty of 40~\tiu, based on the shape of the
  $\chi^2$ curve around the minima of the `no roughness' and `low roughness'
  models (see Figure~\ref{fig:chisq}). The uncertainty on the $p_V$ values are
  derived using the 10\% relative error on $D_\vee$. Table~\ref{tab:TPM}
  summarizes our results.\\

\begin{table}
\centering 
\begin{tabular}{cccccc}
\hline
 {{\small Roughness model}} & {{\small Reduced $\chi^2$}}&{$\Gamma$}&{D$_\vee$}& {$p_V$} \\ 
                   &                      &(\tiu)& (km) &\\
\hline
No roughness & 3.3& $133\pm40$ & {$247\pm25$}&  {$0.12\pm0.02$}\\
Low roughness& 3.2& $114\pm40$ & {$238\pm24$}&  {$0.13\pm0.03$}        \\
\hline    
\end{tabular} 
\caption[Thermal properties]{Results of the determination of physical properties of the 
  asteroid (16) Psyche, using the TPM.
  $\Gamma$ is the thermal inertia, $D_\vee$ is the spherical volume 
  equivalent diameter, and $p_V$ is the geometric visible albedo. The errors represent an estimation of the dominant contribution of the model systematics, as explained in Section~\ref{s:results}.  
}
\label{tab:TPM}
\end{table}
%
%
%
%

\section{Discussion}
\label{s:discussion}
\subsection{Size and albedo} 
  The best-fit values of $D_\vee$ obtained from our TPM analysis of MIDI
  data, $247\pm25$~km and $238\pm24$~km, presents an offset of about 30~km 
  with the nominal value ($211\pm21$~km). Even though these values are in agreement within the error bars, our TPM results seem to favor a larger diameter for (16) Psyche.      
  We remind that the
  condition of convexity, imposed by the lightcurve inversion technique
  \citep[see][]{2001Icar..153...24K}, may introduce such a systematic
  bias on the size determination when large concavities are present on the
  asteroid surface. In this case, the volume-equivalent diameter obtained when
  using a convex shape will overestimate the `true' volume and then the size
  of the asteroid. In our preceding analysis of (41) Daphne with MIDI \citep{2011Icar..215...47M}, we faced a
    similar situation. The volume-equivalent diameter derived
    with a convex shape model was overestimated, and the use of a more
    detailed shape model, including concavities \citep{Carry2009},
    solved this discrepancy.
    Although the diameter values of 247\,km and 238~km we derived here are close to the
    average diameter of all reported estimates for Psyche
    \citep[$247\pm19$\,km, see][]{2012P&SS...73...98C}, we expect the real value to be
    somehow smaller.\\

  \indent Using our TPM diameter estimation, we derived a geometric visible albedo
  value in the range between 0.12 and 0.13. As expected from the IRAS and
  AKARI size estimates, our estimate lies between the values derived from
  those surveys, namely $0.120\pm0.004$ and $0.18\pm0.01$, respectively. Those
  values are not identical within uncertainties. This discrepancy is probably
  due to the fact that we are comparing our volume equivalent diameter
  estimate with instantaneous area equivalent diameters measured at different
  epochs. Nevertheless, those albedo values are in good agreement with the 
  M-type taxonomic type of 
  (16) Psyche (by definition, M-types have an albedo between 0.075 and 0.30). \\ 


\subsection{Thermal properties} 
  As detailed in the introduction, infrared interferometry can spatially probe
  the asteroid surface temperature distribution in different directions at high
  angular resolution. Used in combination with infrared radiometry, it can help
  to remove the degeneracy existing in our contraints of thermal inertia and
  surface roughness and this at one single epoch, as already shown in
  \citet{2011Icar..215...47M}.  

  We obtained good constraints on the
  determination of macroscopic roughness of (16)
  Psyche. Both no- and low- roughness models appears equally good in
  terms of our best-fit estimator.   
  We estimate that the corresponding mean surface slope, as defined by
  \citet{1984Icar...59...41H}, should be lower than 10$^\circ$ for (16)
  Psyche. A high macroscopic 
  roughness is discarded. We remind that the roughness is at scales ranging
  from several centimeters to a fraction of the length of a facet, the latter
  being of the order of 10 kilometer.
  Interestingly, radar data reported by \citet{2008Icar..195..184S} give a very low polarization ratio of $0.06\pm0.02$ for (16) Psyche, which indicates a smooth
  surface without significant radar-wavelength-scale surface roughness. Nevertheless, this agreement has to be considered with caution since the macroscopic surface roughness probed by infrared interferometry may be at a scale different than that constrained by radar observations.

\begin{figure}
 \centering
 \includegraphics{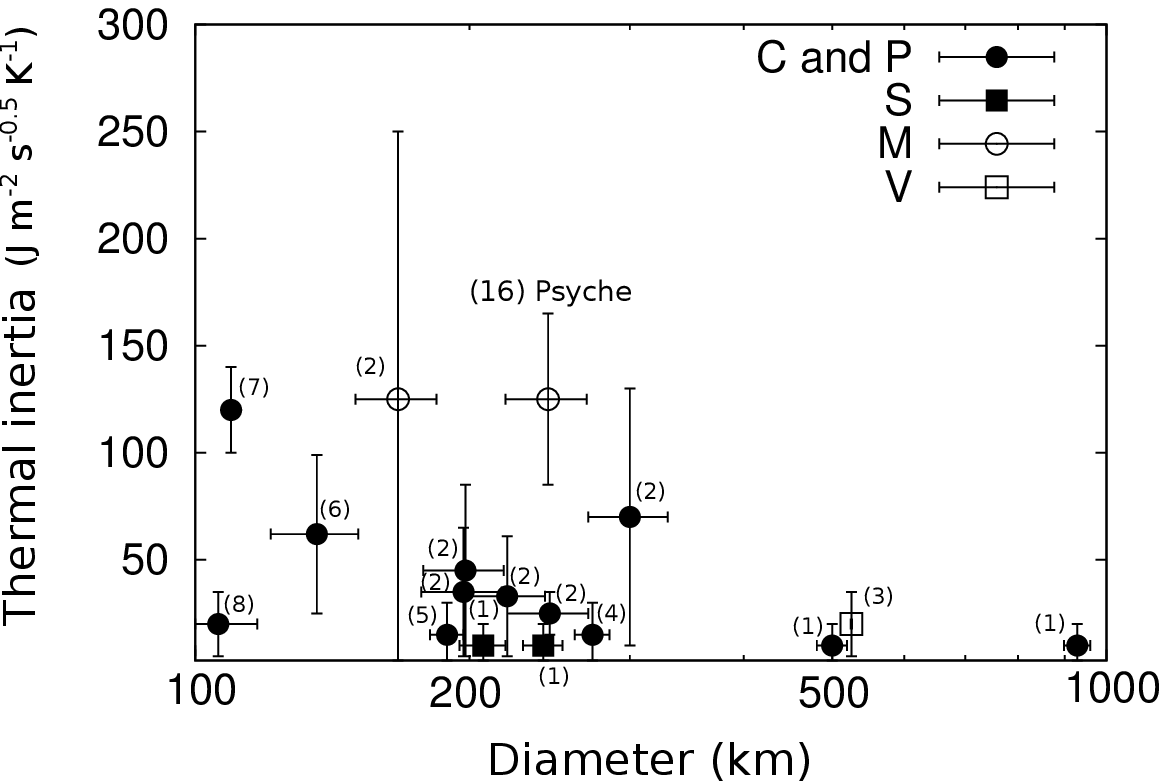}
 \caption[Thermal inertia vs asteroid diameter]{
   Compilation of the existing thermal inertia measurements of main-belt asteroids larger than 100~km in diameter. We included our new measurement for (16) Psyche. We indicated the taxonomic type (C-complex, S-complex and V-type) of each asteroid following \citet{2002Icar..158..146B,2009Icar..202..160D}. We separated the X-complex into M-type and P-type asteroids, following the Tholen classification, to emphasize the higher thermal inertia of M-type asteroids. We also mention the corresponding references for each measurement: (1) \citet{1998A&A...338..340M}, (2) \citet{2012Icar..221.1130M}, (3)
   \citet{2012A&A...539A.154L}, (4) \citet{2004A&A...418..347M}, (5)
   \citet{2011Icar..215...47M}, (6)
   \citet{2010Icar..205..505M}, (7) \citet{2009P&SS...57..259D}, (8) \citet{2012P&SS...66..192O}.}   
 \label{fig:diam_TI}
\end{figure}

  In addition, our TPM analysis indicates that (16) Psyche has a thermal
  inertia value lying between $114\pm40$ and $120\pm40$~\tiu. This is significantly larger
  than the thermal 
  inertia values generally measured on main-belt asteroids larger than 100~km in
  diameter (see Figure~\ref{fig:diam_TI}). Indeed, large main-belt asteroids such as (1) Ceres, (2) Pallas,
  (3) Juno, (4) Vesta, (21) Lutetia, (41) Daphne, (65) Cybele, or (532)
  Herculina present a very low thermal inertia, between 5 and 30~\tiu
  \citep{1998A&A...338..340M,2004A&A...418..347M,2010A&A...516A..74L,2011Icar..215...47M}. \citet{2010Icar..205..505M}
  also measured a low thermal inertia of\\ 
  $20\pm15$~\tiu for the large binary
  Trojan (617) Patroclus. More recently, \citet{2012Icar..221.1130M} also
  found thermal inertia potentially lower than 100~\tiu~for 200~km-class
  main-belt asteroids, except for the M-type asteroid (22) Kalliope, which could exhibit a higher value 
  ($\Gamma=5-250$~\tiu). One exception is the asteroid (694) Ekard for which \citet{2009P&SS...57..259D} determined a thermal inertia value around
    100-140~\tiu. However, the TPM fit to the IRAS data for this
    asteroid was the worst among those studied in that work,
    indicating that the thermal inertia value for (694)
    Ekard might be less accurate compared to the other values derived for 
    asteroids with sizes larger than 100 km.
	All those
  measurements imply that the surfaces of 
  those large bodies are likely covered by a thick layer of fine-grained dust regolith. This is
  expected for such large bodies that can gravitationally retain on their
  surface loose material like thin dust produced from impact ejecta. On the
  other hand, smaller asteroids of a few km or tens of km in size have
  lower gravity and are expected to retain less regolith from impacts.\\   
  In
  general, existing measurements shows a decrease in the thermal inertia value
  with increasing asteroid diameter  \citep[see
    e.g.,][]{2007Icar..190..236D,2009P&SS...57..259D}. This suggests that the thermal inertia of asteroids is mainly controlled by the
  dusty nature of the asteroid surface rather than other physical parameters
  associated to the nature of the material itself. This especially concerns
  thermal conductivity as thermal inertia only varies as the square root of
  the conductivity. 
  Since (16) Psyche is a large main-belt asteroid, its surface should be also
  covered by such a thick and thermally insulating dust regolith, which would
  imply a low thermal inertia. Therefore, the high thermal inertia value we
  measured strongly suggests a significant difference in terms of composition,
  so that it would sufficiently increase the thermal conductivity of its
  surface and then its thermal inertia. This may also explain the potentially high thermal inertia (5-250~\tiu) measured for the large M-type asteroid (22) Kalliope \citep{2012Icar..221.1130M}.  
  As stated in the introduction, metal is an excellent thermal conductor,
  potentially leading to an enhanced thermal inertia. In this context, 
  \citet{2010Icar..208..449O} measured the thermal conductivity of a sample of
  meteorites including two ordinary chondrites, one enstatite chondrite, two
  carbonaceous chondrites and one iron meteorite. They showed that the thermal
  conductivity, at low temperatures (5 to 300~K), of iron meteorites is much
  higher, by about one order of magnitude, than the one of stony meteorites,
  especially the ordinary and carbonaceous chondrites. Interestingly, one
  order of magnitude difference in thermal conductivity corresponds to a
  factor 3 to 4 in thermal inertia, which is roughly the discrepancy between
  our thermal inertia measurement for (16) Psyche, around 120~\tiu, and
  the average thermal inertia of other large main-belt asteroids, around 30~\tiu. 
As a consequence, a metallic surface composition appears as a realistic
explanation for the high thermal inertia we measured on the surface of (16)
Psyche. Our results thus constitute a new evidence of the metal-rich
composition of (16) Psyche, and confirm the previous radar studies on this
object \citep[e.g.,][]{2010Icar..208..221S}.\\ 

  We show in Fig.~\ref{fig:diam_TI} all the existing thermal inertia
  measurements of main-belt asteroids larger than 100~km in diameter, including our measurement for (16) Psyche. We can see that the two M-type asteroids of the list, (16) Psyche and potentially (22) Kalliope,
  somehow sticks out from the thermal inertia range of large MBAs. This difference is even noticeable with P-type asteroids, which show thermal inertia values comparable to the one of C-complex and S-complex asteroids. This is in agreement with the expectation that P-type asteroids have surface materials rich in carbon and/or organics \citep{1985Icar...64..503V,2004LPI....35.1616H}, which are less thermally conductive. 
  Figure~\ref{fig:diam_TI} thus illustrates how the relation between
  size and thermal inertia, highlighted for instance in
  \citet{2009P&SS...57..259D}, can be modified when asteroids with similar sizes but different
  compositions, especially metallic, are considered.  \\  
\section{Summary}

  We have obtained the first successful interferometric observations of the
  M-type asteroid (16) Psyche using the MIDI instrument and the 16m-long
  baseline E0-G0 of the VLTI.\\ 
  Following the work of \citet{2011Icar..215...47M}, we applied our
  thermophysical model (TPM) to the MIDI observations of (16) Psyche 
  to derive its size and the thermal properties of its surface.
  Using the convex
  shape model of \citet{2002Icar..159..369K}, our TPM results indicate that
  Psyche has a volume equivalent diameter between $238\pm24$ and $247\pm25$~km, depending on
  the assumed surface roughness.\\  

  \indent Our analysis also showed that a low macroscopic surface roughness is
  clearly favored by our interferometric observations, and that `high
  roughness' models are discarded.  With such a 
  constraint on the macroscopic roughness, the TPM results indicate a
  high thermal inertia for (16) Psyche, of 130~\tiu (`no roughness') or
  114~\tiu (`low roughness'), with a total uncertainty estimated at
  40~\tiu. This is one of the highest thermal inertia ever 
  measured for a 200~km-class asteroid. This is in clear contradiction with
  previous results indicating that the surface of asteroids with sizes larger
  than 100 km have a low thermal inertia. As metal is an excellent thermal
  conductor, we expect this high thermal inertia to be another evidence of the
  metallic composition of (16) Psyche, as previously inferred from radar
  studies. This reinforces the hypothesis of (16) Psyche as originated from
  the fragmentation of the iron core of a differentiated parent body or more
  likely, considering its size, from the collision and aggregation of several
  objects, with at least one of them being purely metallic.\\  

\subsection*{Acknowledgments}
We would like to thank the referees for their comments and suggestions that helped to improve significantly this manuscript.
We would also like to thank the staff and the Science Archive Operation of
the European Southern Observatory (ESO) for their support in the
acquisition of the data. The development of the asteroid thermophysical model used in this work was partially supported by the project 11-BS56-008 (Shocks) of the Agence National de la Recherche (ANR).
  This research used the Miriade VO tool \citep{2008LPICo1405.8374B} developed at IMCCE.


\bibliographystyle{elsarticle-harv}
\bibliography{biblio_16psyche}

\end{document}